\documentclass[twocolumn]{autart}
\makeatletter
\def\@journal{}
\makeatother

\usepackage{graphicx} 
\usepackage[utf8]{inputenc}
\usepackage{amsmath,amssymb,amsfonts}
\usepackage{algorithmic}
\usepackage{cite}
\usepackage{newtxmath}
\usepackage{color}
\usepackage{url}

\newtheorem{defin}{Definition}

\begin{document}

\begin{frontmatter}

\title{Stability analysis of nonlinear stochastic flexibility function in smart energy systems} 


\author[]{Seyed Shahabaldin Tohidi}\ead{sshto@dtu.dk},    
\author[]{Tobias K. S. Ritschel}\ead{tobk@dtu.dk},  
\author[]{Georgios Tsaousoglou}\ead{geots@dtu.dk},
\author[]{Uffe H{\o}gsbro Thygesen}\ead{uhth@dtu.dk},
\author[]{Henrik Madsen}\ead{hmad@dtu.dk}

\address[]{Department of Applied  Mathematics and Computer Science, Technical University of Denmark, DK-2800 Kgs. Lyngby, Denmark}  


\begin{abstract}   
Demand-side management provides a great potential for improving the efficiency and reliability of energy systems. This requires a mechanism to connect the market level and the demand side. The flexibility function is a novel approach that bridges the gap between the markets and the dynamics of physical assets at the lower levels of the energy systems and activates demand-side flexibility with the purpose of decision-making as well as for offering a new framework for balancing and grid services. Employing this function as a key for many decision-making and control algorithms reveals that a mathematically rigorous stability analysis is required for it. In this paper, we investigate the stability properties of two nonlinear flexibility functions, as a dynamic mapping between electricity price and power consumption. Specifically, we analyze the stability of a deterministic flexibility function and an It\^{o} stochastic flexibility function. Simulation results are also provided to demonstrate the dynamics of the flexibility functions and to show that the analytical results hold.

\end{abstract}

\end{frontmatter}

\section{Introduction}

Decarbonization of the energy system forms an immense priority towards the reduction of the rate of global warming \cite{WMO22}. Thereupon, energy systems worldwide are now subject to ambitious goals regarding penetration levels of renewable energy sources (RES). 
Such developments, however, challenge the traditional operations of power systems, where the supply side is controlled to meet the inflexible demand and call for control capabilities on the demand side for the latter to meet the intermittent renewable supply \cite{Zhu16}, \cite{Zhang15}, \cite{Mo20}. 

This paradigm shift brings about the development of control and communication capabilities toward harvesting the innate flexibility of existing electricity loads. This has led to an outburst of research studies for demand-side energy management, spanning different types of loads (e.g. residential \cite{pallonetto2020assessment}, industrial \cite{shoreh2016survey}, cross-sector \cite{gjorgievski2021potential}) and methodological approaches (e.g. optimization \cite{jordehi2019optimisation}, control \cite{farrokhifar2021model}, artificial intelligence \cite{vazquez2019reinforcement}). 
A characteristic example refers to the management of thermostatically controlled loads to offer load-shifting services by leveraging the natural thermal storage capability of the building envelope \cite{Pean19}, \cite{vrettos2016experimental}, \cite{Coffman23}.

The multitude of small, demand-side resources along with their different characteristics and their heterogeneous control methods constitutes a major barrier towards their coordination and integration into hierarchical control \cite{yamashita2020review}, \cite{DeZotti18}, \cite{DeZotti19} or market \cite{tsaousoglou2022market}, \cite{pandey2021hierarchical}, \cite{tsaousoglou2021demand} frameworks.
A prominent approach towards achieving scalable coordination frameworks for flexible electricity loads refers to the so-called ``transactive control'' paradigm \cite{hao2016transactive}, which constitutes a market-based hierarchical control method, where a resource's controller receives a set of prices (communicated by an upstream controller) and adapts its consumption profile as a response to these prices considering also its internal state and characteristics.

From a practical perspective, however, it remains an open challenge to integrate heterogeneous resource-level controllers into an upstream coordination framework that also considers uncertainty and schedules operations for times ahead. A solution to this problem is to capture a resource's (e.g. building's) flexibility capabilities into a generic grey-box model \cite{Tohidi22} that can be used seamlessly in upstream control or market frameworks \cite{Rune18}. Such a model is instantiated in \cite{Rune20} as a set of stochastic differential equations that provide a nonlinear mapping between a set of electricity prices (for a look-ahead horizon) and a resource's power consumption profile that comes as a response to those prices, and coined as a resource's ``Flexibility Function''. As further showcased in \cite{Rune20}, \cite{Domink20}, and \cite{Thilker12}, by simply tuning the parameters of the Flexibility Function, we can use it to capture the price-consumption behavior of different types of loads and their respective controllers. This offers great potential towards a seamless integration of distributed flexibility capabilities and it is currently considered as a cornerstone of minimum interoperability mechanisms for a smart operating system under development by the Danish power system operator.

Stability analysis of the energy system and controller design are investigated in many research articles. For instance, the stability of the energy system considering dynamic-pricing controllers and distributed control architecture are discussed in \cite{Stegink16} and \cite{Zhang15}, respectively. 
To enable the adoption of the Flexibility Function as a standard in real systems, it is of prominent importance that the model that instantiates a resource's Flexibility Function is theoretically proven to be stable, i.e., it always provides a meaningful output of the resource's expected consumption profile once provided with a set of prices. 
This motivates the main contribution of this paper, which is to identify the region of the Flexibility Function's parameters, within which the model remains stable. Specifically, the stability of a deterministic Flexibility Function and an It\^{o} \cite{Uffe23} stochastic Flexibility Function is analyzed. This provides a solid theoretical foundation for the model's use in the critical domain of power systems, while also providing a practical tool for ensuring that the tuning of the Flexibility Function upon modeling any particular resource stays within a meaningful and well-behaved region. The other contribution of this study is to translate the flexibility function into an input-state-output form, which allows it to fit into the standard paradigm for systems theory.  To the best of our knowledge, such a result is not available in the prior literature. This approach makes use of concepts such as equilibrium points, stability, and boundedness, which in turn simplifies the analysis of flexibility dynamics.

This paper is organized as follows: Section 2 gives an overview of the Flexibility Function. Sections 3 and 4 provide the stability analysis of the deterministic and stochastic Flexibility Functions, respectively. Section 5 demonstrates the simulation results to demonstrate stability analysis results.
Finally, a summary is given in Section 6.

\subsection{Notation}
In this paper, $\mathbb{R}$ denotes the set of real numbers. Elements of a set can be provided using $\{.\}$. Closed interval is shown as $[ a, b]=\{x\in \mathbb{R}| a\leq x \leq b  \}$. The symbol $/$ is used for set subtraction. The symbol $\wedge$ stands for the logical and, i.e., $a\wedge b$ is true if both $a$ and $b$ are true. $\Pr(.)$ and $E[.]$ denote probability and expected value, respectively. Also, the normal distribution with a mean of $\mu$ and a standard deviation of $\sigma$ is
denoted by $N(\mu, \sigma )$. If $X$ has normal distribution, we write $X\sim N(\mu, \sigma)$. Moreover, $C^2$ stands for twice continuously differentiable, and $tr(.)$ refers to the trace operation. $\vmathbb{1}_{q}$ is the indicator function, i.e., it is equal to one only if the logic $q$ is true. Finally, $||.||$ is the $L_2$ norm and $|c|$ is the absolute value of $c\in \mathbb{R}$.


\section{Flexibility Function}\label{sec:FF}

Flexibility is the capability of adjusting the energy demand profile according to climate conditions, grid requirements, and comfort constraints. Energy systems have the flexibility potential. Exploiting this potential is a key task for energy management and load scheduling in modern energy systems. Lack of predictability on the demand side raises the need for a systematic way to represent the flexibility potential and predict it.
Flexibility function is a dynamical system that introduces a linkage between demand, state of charge, and price \cite{Rune18}. 

The nonlinear dynamics of the price-demand relationship, as proposed in \cite{Rune20}, is in the following stochastic differential equation form. The dynamics of the state of charge of flexibility function is
\begin{align}
   d{\mathcal{X}}_t&=\frac{1}{C}\Delta D_t dt + \mathcal{X}_t(1-\mathcal{X}_t)\sigma_xdW_t,\label{eq:1} 
\end{align}
where $\mathcal{X}\in [0,\ 1]\subset \mathbb{R}$ is the state of charge, $C \in \mathbb{R}^+$ is the capacity of flexible energy, $\sigma_X$ represents process noise intensity, and $W$ is a Wiener process. It is seen that positive $\Delta D_t$ leads to an increase in $\mathcal{X}_t$ and ultimately reduces the diffusion. On the contrary, if $\Delta D_t$ is negative, $\mathcal{X}_t$ reduces and ultimately reduces the diffusion. Furthermore, $\Delta D\in [-1,\ 1]\subset \mathbb{R}$ is defined as the following function, 
\begin{align}
    \Delta D_t = \begin{cases}
\delta_t \lambda (1-B_t), & \text{ if}\ \  \delta_t\geq 0,\\
\delta_t \lambda B_t, & \text{ if}\ \  \delta_t < 0,
\end{cases}
\end{align}
where $B_t\in [0,\ 1]\subset \mathbb{R}$ is the baseline demand, $\lambda\in (0,\ 1)\subset \mathbb{R}$ is the portion of the demand and $\delta_t$ is the demand change. If $\delta_t$ is positive, i.e., the state of charge is increasing, then, $\Delta D_t$ is equal to $\delta_t$ multiplied by a portion of the difference between full demand and the baseline demand $(1-B_t)$. Otherwise, $\Delta D_t$ is equal to $\delta_t$ multiplied by a portion of the difference between the null demand and the baseline demand $(B_t)$. $\Delta D_t$ can be interpreted as the deviation of the predicted demand from the baseline demand as
\begin{align}
    D_t = B_t +\Delta D_t,
\end{align}
where $D_t[0,\ 1]\subset \mathbb{R}$ is the predicted demand that can be considered as the flexibility function output.


The nonlinear function defining 
$\delta_t$ is a logistic function of $\mathcal{X}_t$ and the price $(u_t)$ as:
\begin{align}
    \ell(z_t) &= -1 + \frac{2}{1+e^{-kz_t}},
\end{align}
where $z_t \equiv f(\mathcal{X}_t)+g(u_t)$, where $f(.)\in [-1,\ 1]$ is a polynomial function between the state of charge and energy demand and $g(.)\in [-1,\ 1]$ is a summation of I-splines representing the relation between the energy price and energy demand \cite{Ramsay1988} as 
\begin{align}
g(u_t) &= \beta_1 I_{s,1}(u_t) + ... + \beta_7 I_{s,7}(u_t),\label{eq:3y}\\
f(\mathcal{X}_t) &= \left(1-2\mathcal{X}_t+\alpha_1\left(1-(2\mathcal{X}_t-1)^2\right)\right)\notag\\ &\times \left(\alpha_2+\alpha_3(2\mathcal{X}_t-1)^2+\alpha_4(2\mathcal{X}_t-1)^6\right),\label{eq:4y}
\end{align}
where $I_{s,1}$, ..., $I_{s,7}$ are I-spline functions , $\beta_1, ..., \beta_7$, $\alpha_1, ..., \alpha_4$, and $k$ are parameters to be identified. These parameters can be identified using different approaches to maximize the likelihood between the final flexibility function and the data. The following two remarks reveal some useful aspects of determining $f(.)$ and $g(.)$.

\begin{figure}
\centering
\includegraphics[width=0.45\textwidth]{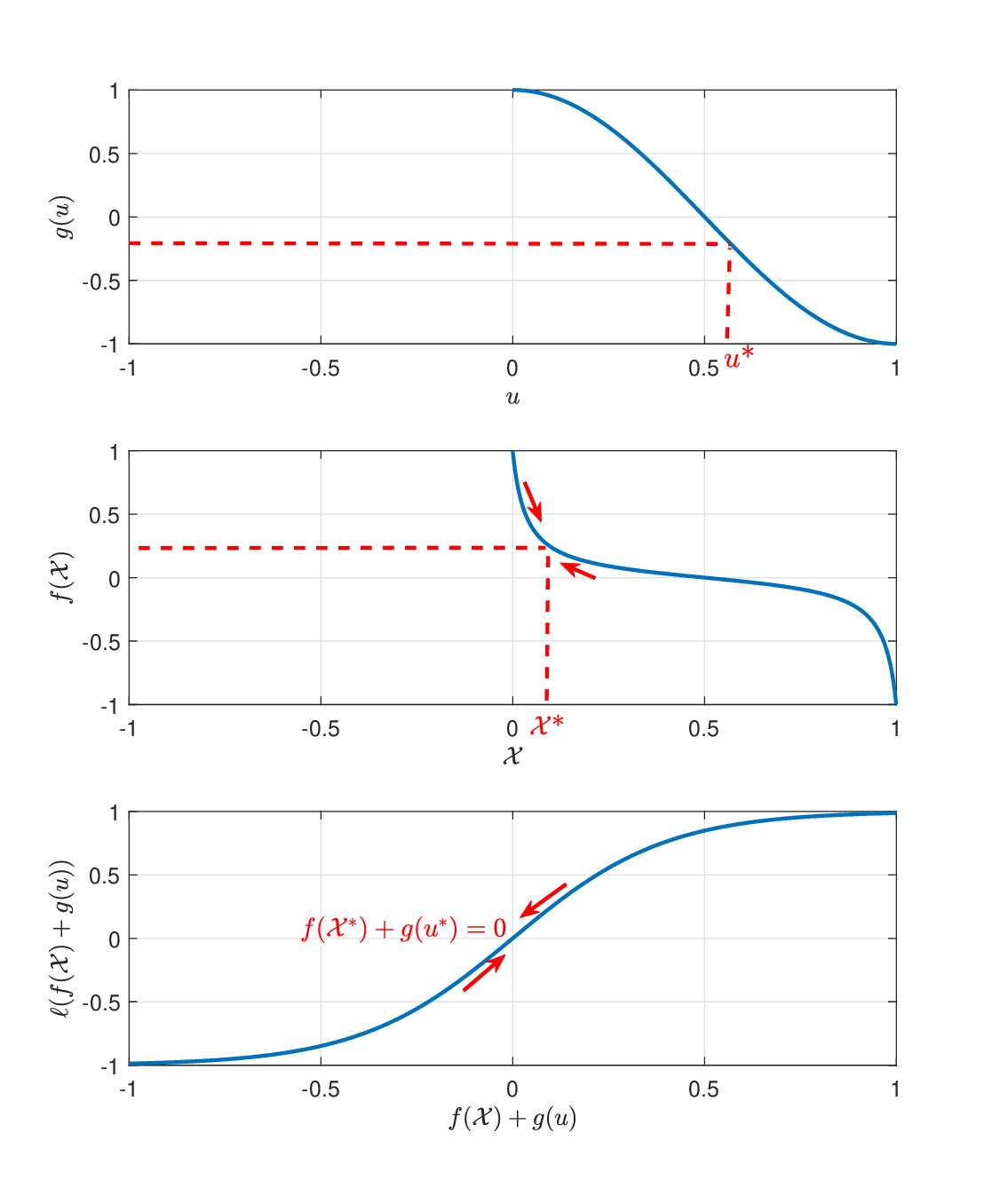}
\caption{Schematic of nonlinear functions of the flexibility function for the stability analysis. The functions $f$ and $g$ are selected to demonstrate the analysis.}
\label{fig:stable}
\end{figure}

\begin{rem} \label{rem1x}
    The following three concepts form the basis of the flexibility function dynamics: 1) high prices lead to a reduction in demand, while low prices tend to increase demand, 2) more stored energy leads to a reduction in demand, while less stored energy tends to increase demand, and 3) the energy storage capacity is finite. As a result of the first two concepts, the functions $f(.)$ and $g(.)$ are determined such that they are monotonically decreasing. Furthermore, the function $\ell(.)$, mapping $f(.) + g(.)$ to the interval $(0,1)$ is designed to be monotonically increasing, which represents the effect of having low/high price and low/high state of charge at the same time (see Figure \ref{fig:stable}).
\end{rem}

\begin{rem} \label{rem2x}
    The functions $f(.)$ and $g(.)$ are selected in a way that $f(1)+g(u)\leq 0$ and $f(0)+g(u)\geq 0$ for all $u\in [0,\ 1]$. Considering that $u=1$ minimizes $g(u)$, and $u=0$ maximizes it, $f(.)$ and $g(.)$ are chosen such that  $f(1)+g(0)= 0$ and $f(0)+g(1)= 0$ (see Figure \ref{fig:stable}).
\end{rem}

In the following, we first disregard the stochastic terms in (\ref{eq:1}) and investigate the stability of the deterministic flexibility function. Then, we analyze the stochastic dynamics of (\ref{eq:1}) as a general model for demand-response dynamics.


\section{Stability Analysis of the Deterministic Flexibility Function}

Disregarding the diffusion part of the flexibility function provides a clarified overview of the demand-response behavior. This simplified ordinary differential equation helps us understand the core of its dynamics and can be viewed as an introduction to a more complex stochastic flexibility function. Other studies like \cite{Amos24} have utilized this simplified flexibility function as a price-demand relationship. To this end, the flexibility function can be written as
\begin{align}
d{\mathcal{X}}_t/dt&=\frac{1}{C}\Delta D_t ,\label{eq:1x}\\
\Delta D_t &= \begin{cases}
\delta_t \lambda (1-B_t), & \text{ if}\ \  \delta_t\geq 0,\\
\delta_t \lambda B_t, & \text{ if}\ \  \delta_t < 0,
\end{cases},\label{eq:2x}\\
D_t&=B_t+\Delta D_t,\label{eq:3x}\\
\delta_t&=\ell(z_t),\label{eq:4x}\\
z_t&=f(\mathcal{X}_t)+g(u_t),\label{eq:4xy}
\end{align}
for the further analysis. In the following, we first investigate the flexibility function by finding equilibrium points of it, and then analyze its stability using Lyapunov stability analysis.
\subsection{Set invariance property}
In this subsection, we show that the flexibility function (\ref{eq:1x})-(\ref{eq:4xy}) has the property to keep the state of charge in the range $[0,\ 1]$. Specifically, we validate the flexibility function at $\mathcal{X}_0=1$ and $\mathcal{X}_0=0$. 

Suppose that the price is $\bar{u}$, the baseline demand is $\bar{B}\in (0,\ 1)$, and $\mathcal{X}_0=1$. Then, using the properties of $f(.)$ and $g(.)$, introduced in Remark \ref{rem1x} and \ref{rem2x}, it implies that $f(\mathcal{X}_0)=-1$, which leads to $f(\mathcal{X}_0)+g(\bar{u})<0$. Consequently, from (\ref{eq:4xy}), (\ref{eq:4x}) and (\ref{eq:2x}), it can be obtained that $z<0$, $\delta<0$, and $\Delta D<0$. This shows that $d{\mathcal{X}}_t/dt<0$, and proves that $\mathcal{X}$ will not leave the set $[0,\ 1]$. The same argument holds for the price $\bar{u}$, the baseline demand $\bar{B}\in (0,\ 1)$, and $\mathcal{X}_0=0$. One can show that $d{\mathcal{X}}_t/dt>0$, which proves that $\mathcal{X}$ will not leave the set $[0,\ 1]$.

\subsection{Equilibrium points}\label{sec:EP}

Equilibrium points are crucial in determining the stability and convergence properties of a dynamic system. They play a significant role in stability theorems by helping to investigate the behavior of the system in the vicinity of the equilibrium points. Therefore, it is important to understand the concept of equilibrium points when analyzing the stability of a dynamic system.

The equilibrium points of the system are the points that make the right-hand side of (\ref{eq:1x}) equal to zero, that is, $\Delta D_t=0$. Hence, once demand matches the baseline, the dynamics of the flexibility function are in equilibrium. Considering (\ref{eq:2x}), this happens when $\delta_t=0$ or equivalently, $\ell(f(\mathcal{X}_t)+g(u_t))=0$. Thus, the set consisting the equilibrium point(s), $X^*$, of the system can be defined as 
\begin{align}\label{eq:6cd}
 \Omega = \{ (X^*, u^*)| f({\mathcal{X}})=-g(u^*) \}   
\end{align}
for the specified $u^*$ and $B^*$. It is noted that since there exists a unique $u^*$ corresponding to each $\mathcal{X}$ (see Remark \ref{rem1x} and \ref{rem2x}), we denote the equilibrium points as a pair, $(X^*, u^*)$.

\begin{rem}\label{rem1}
    One may also ask about other points that make the right-hand side of (\ref{eq:1x}) equal to zero. These points are related to the extremum values of the baseline signal. It is seen in (\ref{eq:2x}) that $\Delta D_t$ is equal to zero when $B_t=0$ and $\delta_t<0$, or $B_t=1$ and $\delta_t\geq 0$. Although this proposition may look correct, by looking in more detail it reveals that it is not physically meaningful. When, $B_t=0$, $D_t$ is greater or equal to zero, leading to $\delta_t\geq 0$. This is the same for $B_t=1$. When, $B_t=1$, $D_t$ is less than or equal to $1$, leading to $\delta_t\leq 0$.
\end{rem}


To simplify the stability analysis, one can, without loss of generality, apply a change of variable, and investigate the stability around the origin. The transformed system is written as:
\begin{align}
\frac{d{\mathfrak{X}}_t}{dt}&=\frac{1}{C}\Delta D_t,\label{eq:1xx}\\
\Delta D_t &= \begin{cases}
\delta_t \lambda (1-B_t), & \text{ if}\ \  \delta_t\geq 0,\\
\delta_t \lambda B_t, & \text{ if}\ \  \delta_t < 0,
\end{cases},\label{eq:2xx}\\
D_t&=B_t+\Delta D_t,\label{eq:3xx}\\
\delta_t&=\ell(z_t),\label{eq:4xx}\\
z_t&=f(\mathfrak{X}_t+X^*)+g(u_t),\label{eq:4xyx}
\end{align}
where $\mathfrak{X}=\mathcal{X}-X^*$, $X^*\in \Omega$. The equilibrium point of the transformed system is at 0. In the sequel, we analyze the stability of the dynamics (\ref{eq:1xx})-(\ref{eq:4xyx}).

\subsection{Stability analysis of the equilibrium points}
This section is dedicated to the stability analysis of the equilibrium points of the flexibility function, disregarding the stochasticity, using Lyapunov stability theorem \cite{Khalil2002}.

\textbf{Theorem 1}\label{thm1}
    \textit{Consider the transformed flexibility function (\ref{eq:1xx})-(\ref{eq:4xyx}) and let $\mathbb{D}$ be a domain containing the origin. Then given any initial condition in $\mathbb{D}$, and any $u^*\in [0,\ 1]$ and $B^*\in [0,\ 1]$, the equilibrium point of (\ref{eq:1xx})-(\ref{eq:4xyx}) is asymptotically stable.
    }
    
\textbf{Proof}
    Consider the Lyapunov function candidate $V(\mathfrak{X}_t)=\frac{1}{2}\mathfrak{X}_t^2$.
If one can show that $\frac{d}{dt}V(\mathfrak{X}_t)< 0$ for $\mathfrak{X}_t\neq 0 $, then the stability of the equilibrium point is guaranteed. The time derivative of $V$ along the trajectories of $\mathfrak{X}_t$ can be calculated as
\begin{align}
\frac{d}{dt}V(\mathfrak{X}_t)=\mathfrak{X}_t\frac{d}{dt}\mathfrak{X}_t= \frac{1}{C}\Delta D_t\mathfrak{X}_t.\label{eq:7xx}
\end{align}
Considering $u_t=u^*$ and $B_t=B^*$, and substituting (\ref{eq:2xx}), (\ref{eq:7xx}) can be rewritten as
\begin{align}
\frac{d}{dt}V(\mathfrak{X}_t)&= \begin{cases}
\frac{1}{C}\mathfrak{X}_t\delta_t \lambda (1-B_t), & \text{ if}\ \  \delta_t\geq 0,\\
\frac{1}{C}\mathfrak{X}_t\delta_t \lambda B_t, & \text{ if}\ \  \delta_t < 0,
\end{cases}.
\label{eq:8xx}
\end{align}
By utilizing Remark \ref{rem1x} and the monotonicity of $f(.)$, $f(\mathcal{X})+g(u^*)>0$ implies $\mathcal{X}<X^*$ and $f(\mathcal{X})+g(u^*)<0$ implies $\mathcal{X}>X^*$, for $X^*\in \Omega$. The following two cases should be considered, 1) $\delta_t>0$, $X^*\in \Omega$, and 2) $\delta_t<0$,  $X^*\in \Omega$:\\
\textbf{Case 1:}\\ 
This case considers $\delta_t>0$ which leads to $\Delta D_t>0 $. Thus, $\mathfrak{X}_t$ increases while $\mathfrak{X}_t<0$. Therefore, $\mathfrak{X}_t$ converges to $0$, that is, $\mathcal{X}_t$ reaches $X^*$.\\
\textbf{Case 2:}\\ 
This case considers $\delta_t<0$ which leads to $\Delta D_t<0$. Thus, $\mathfrak{X}_t$ decreases while $\mathfrak{X}_t>0$. Therefore, $\mathfrak{X}_t$ converges to $0$, that is, $\mathcal{X}_t$ reaches $X^*$.

Considering the above cases, when $\delta_t>0$, $\mathfrak{X}_t$ is negative and when $\delta_t<0$, $\mathfrak{X}_t$ is positive. Figure 1 demonstrates the above cases. Therefore, with the above-mentioned results and (\ref{eq:8xx}), it can be implied that $\frac{d}{dt}V(\mathfrak{X}_t)< 0$ for all $\mathfrak{X}_t\in \mathbb{D}\setminus \{0\}$, which proves the asymptotic stability of the equilibrium point. It is noted that $\Delta D_t=0$ corresponds to the case when the flexibility dynamics are at the equilibrium, that is, $\mathfrak{X}_t=0$.
\qed

\subsection{Simulation results}
Figure 2 demonstrates the simulation results of the deterministic flexibility function concerning changes in initial state, price, and baseline values. The top panel of this figure illustrates the states of charge considering different initial conditions in the range $[0,\ 1]$, while price and baseline are fixed at $u=0.5$ and $B=0.4$, respectively. It is seen that the state of charge remains stable and the trajectories converge to the equilibrium point of the system for any initial state. The next panel of Figure 2 shows the state of charge for changing prices, while the initial state and baseline are constant values as $\mathcal{X}_0$ and $B=0.4$. It is observed that the state is bounded and converges to steady-state values. Low prices lead to a higher state of charge steady-state while the elevation of prices lowers its steady-state. The bottom panel of Figure 2 depicts the state of charge concerning the baseline signal altering. It is seen that the trajectories converge to the different equilibrium points due to changes in $B$. Increasing the baseline increases the rate of change of state of charge. This fact can also be manifested from (\ref{eq:1xx}).

\begin{figure}
\label{fig:sim1}
\centering
\includegraphics[width=0.5\textwidth]{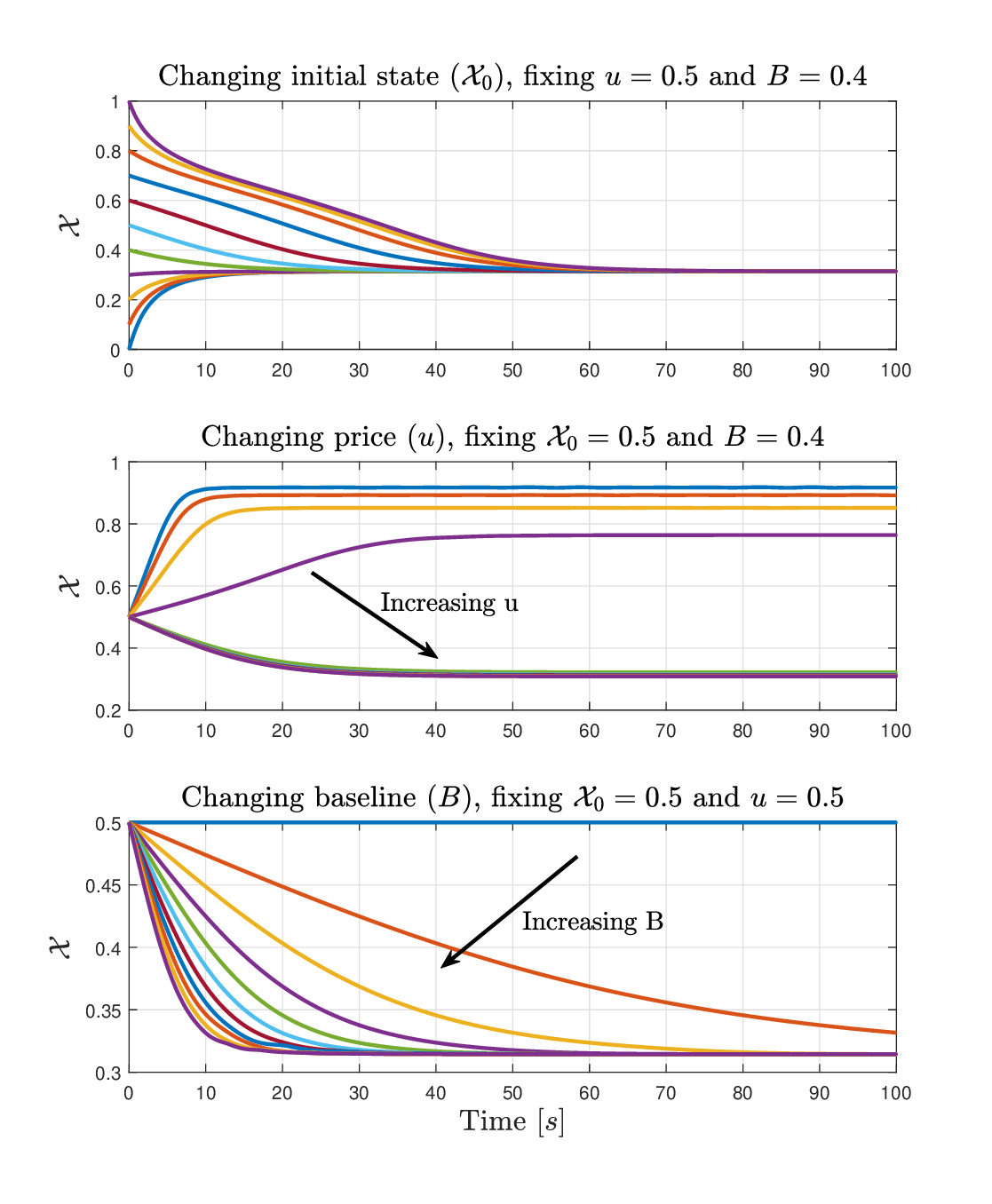}
\caption{Trajectories of state of charge, based on changes in initial state, price, and baseline.}
\end{figure}

\section{Stochastic Stability Analysis}\label{sec:GFF}

Stochastic differential equations are an appropriate approach for modeling physical, biological, and economical dynamical systems with significant uncertainty and where the future values of the states cannot be predicted exactly. The generalized flexibility function is modeled as a stochastic differential equation and consists of both the deterministic part, which is investigated in Section \ref{sec:FF}, and the stochastic part. The stochasticity enables the generalized flexibility function to predict the price-demand relationships realistically. Thus, employing this function in a smart energy Operating System enhances the performance of control and decision-making algorithms. This shows the significance of stability analysis of the stochastic price-demand mapping. 

We first show the importance of the stochastic stability analysis using the following three examples \cite{Sussmann1978}. 

\begin{exmp}
Consider the following bilinear scalar dynamical system
\begin{align}\label{eq7}
\dot{x}(t) = r_1x(t) + r_2 x(t) w(t).
\end{align}
For a deterministic $w(t)$, $r_1+r_2w(t)$ determines the stability properties of this system. Considering the Lyapunov function $x^2/2$, it is simple to show that (\ref{eq7}) is stable if $r_1+r_2w(t)$ is negative for all $t$.
\end{exmp}
\begin{exmp}
For a stochastic input $w(t)$, independent of $x(t)$, expectation of both sides of (\ref{eq7}) can be written as
\begin{align}\label{eq8}
\frac{d}{dt}E[x(t)] &= (r_1 + r_2\omega_t) E[x(t)],
\end{align}
where $\omega_t$ is the mean value of $w(t)$. Using the same discussion as the deterministic input with Lyapunov function $E[x]^2/2$, it can be proven that (\ref{eq8}) is stable if $r_1+r_2\omega_t$ is nonpositive for $\forall t$. However, this is not the case for It\^{o} stochastic differential equations.
\end{exmp}
\begin{exmp}
Consider the It\^{o} stochastic differential equation \cite{Uffe23} 
\begin{align}\label{eq9}
dx(t) =r_1x(t)dt +r_2x(t) dw(t),
\end{align}
where $w(t)$ is a Brownian motion. The explicit solution of (\ref{eq9}) can be found as $x(t) = x(t_0)e^{(r_1-0.5r_2^2)t+r_2w(t)}$ \cite{Uffe23}. This reveals that $x(t)$ is bounded if $r_1-r_2^2/2\leq 0$ and converges to zero if $r_1-r_2^2/2< 0$. 
\end{exmp}
Figure 3 shows the simulation results for the above three examples. This figure shows the difference in the stability properties of seemingly similar systems in Example 1-3. While the systems of Example 1-2 are stable with $r_1=1$ and $r_2=-1.2$, the stochastic differential equation (Example 3) with similar $r_1$ and $r_2$ diverges. On the contrary, with $r_1=1$ and $r_2=2$, the stochastic differential equation of Example 3 converges to zero as time reaches infinity, while systems of Example 1-2 have unstable behaviors. 
\begin{figure}
\label{fig:sim2}
\centering
\includegraphics[width=0.5\textwidth]{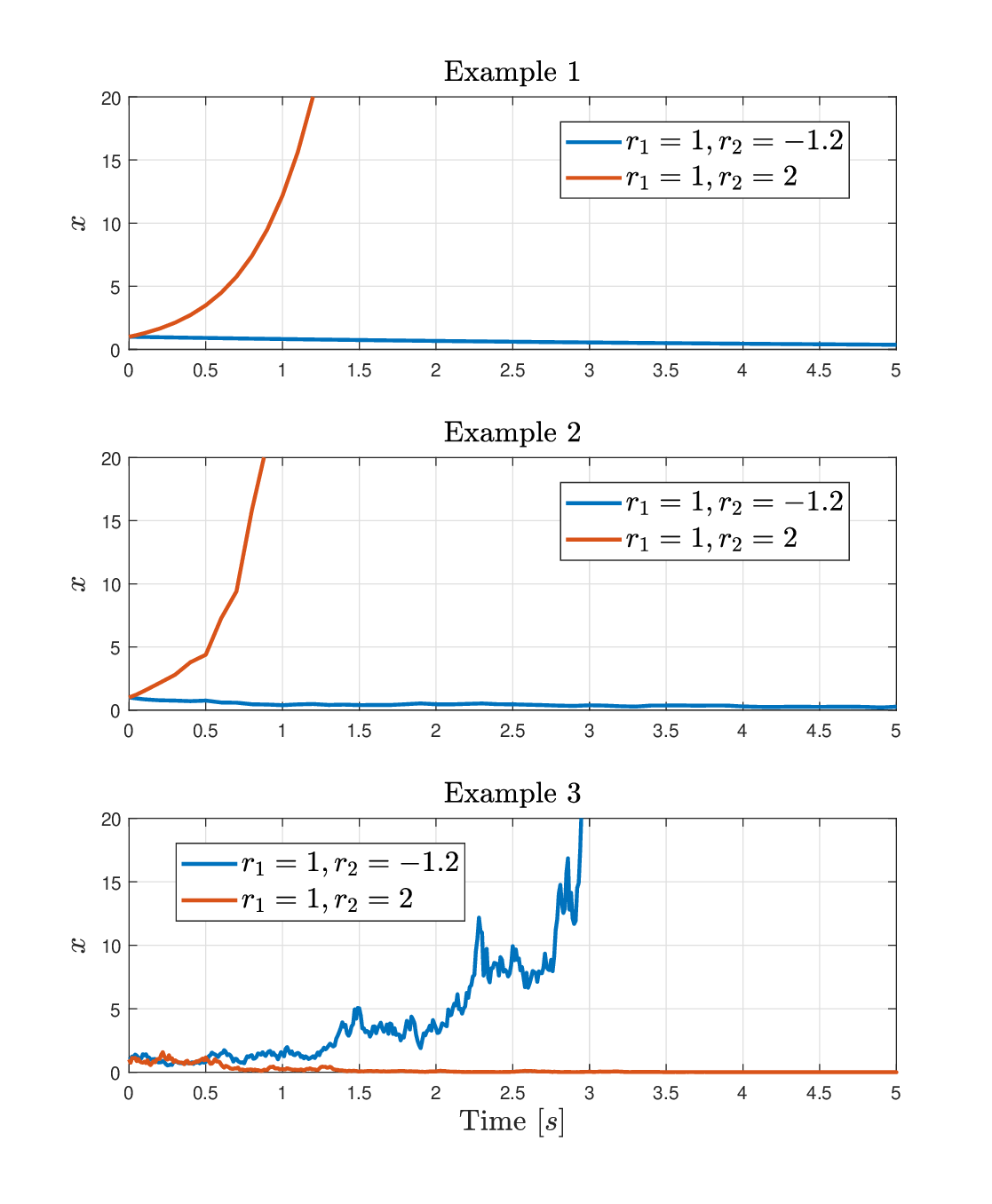}
\caption{State trajectories of the systems in Example 1-3.}
\end{figure}

Before moving toward the stochastic stability analysis of the flexibility function, we should investigate its equilibrium points. Equilibrium points of the stochastic flexibility function (\ref{eq:1})-(\ref{eq:4y}), are the points at which, $\frac{1}{C}(D_t-B_t)=0$ and $ \mathcal{X}_t(1-\mathcal{X}_t)\sigma_x=0$. Since $ \mathcal{X}_t(1-\mathcal{X}_t)\sigma_x=0$ only at $\mathcal{X}=0$ and $\mathcal{X}=1$, the set of equilibrium points of the stochastic flexibility function can be found as
\begin{align}\label{eq:10}
\Omega_s = \{ (\mathcal{X},u^*)| \mathcal{X} \in \{ 0,\ 1 \}\wedge (\mathcal{X},u^*)\in \Omega  \},
\end{align}
where $\Omega$ is defined in (\ref{eq:6cd}). Hence, $\Omega_s=\{(0,\ 1),\ (1,\ 0) \}$. In the sequel, we analyze the stability of the equilibrium points of the stochastic flexibility function.

\subsection{Stochastic stability of flexibility function}
In general, finding an explicit expression for the solution of a system of stochastic differential equations is impossible. However, through a stochastic stability analysis, we can derive a characterization of the behaviors of the solutions. We start the investigation by introducing the main definitions and theorems for the stochastic stability analysis.
\begin{defin} \cite{Kushner1967}
    The solution $X_t$ of the stochastic differential equation $dX_t=f(X_t)dt+g(X_t)dB_t$ with equilibrium $X^*$ is stochastically stable (stable with probability one) if for every $\epsilon >0 $, there exists a $\delta>0$ so that if $||X_{t_0}-X^*||<\delta$, then
    \begin{align}\label{eq:11}
    \lim_{X_{t_0}\rightarrow X^*}\Pr\left(\sup_{t\geq t_0}||X_t-X^*||\geq \epsilon \right) = 0.  
    \end{align}
\end{defin}


\begin{defin} \cite{Uffe97}
    The differential operator $\mathcal{L}$ on function $V(X_t)$ in $C^2$ is defined as
    \begin{align}
    \mathcal{L}V(X_t) = \frac{d}{dX}Vf(X_t)+\frac{1}{2}\text{tr}\left( g(X_t)^T\frac{d^2V(X_t)}{dX^2}g(X_t)\right)
    \end{align}
    where the dynamics of $X_t$ follows the stochastic differential equation $dX_t=f(X_t)dt+g(X_t)dB_t$.
\end{defin}

\textbf{Theorem 2}\label{thm2}\cite{Gard1988}
    \textit{Let $X^*$ be an equilibrium point of the stochastic differential equation $dX_t=f(X_t)dt+g(X_t)dB_t$. Suppose that there exists a function $V$ defined on the domain $\hat{\mathbb{D}}$, containing $X^*$, such that:\\
    1) $V$ is $C^2$ on $\hat{\mathbb{D}}\setminus
    {{X}^*}$,\\
    2) There exist positive definite functions $k_1$ and $k_2$, such that $k_1(|X-X^*|)\leq V(X)\leq k_2(|X-X^*|)$,\\
    3) $\mathcal{L}V(X)\leq 0$ for $|X-X^*|\geq K$ with some constant $K$.\\
    Then, the solution of the stochastic differential equation is stochastically bounded around $X^*$, that is, for each $\epsilon>0$, there exists a $\rho$ such that
    \begin{align}\label{eq:11x}
    \Pr\left(\sup_{t\geq t_0}||X_t-X^*||\leq \epsilon \right) > 1-\rho.  
    \end{align}
}


\textbf{Theorem 3}\label{thm3} \cite{Uffe23}
    \textit{Let $X^*$ be an equilibrium point of the stochastic differential equation $dX_t=f(X_t)dt+g(X_t)dB_t$. Suppose that there exists a function $V$ defined on the domain $\hat{\mathbb{D}}$, containing $X^*$, 
    such that the first and second conditions of Theorem \ref{thm2} are satisfied and $\mathcal{L}V(X)\leq 0$ for $X\in \hat{\mathbb{D}}\setminus
     {{X}^*}$.
    Then, the solution of the stochastic differential equation is stochastically stable.
}

After discussing the importance of stochastic stability analysis for the stochastic differential equation, and providing the required definitions and theorems in this regard, it is time to investigate the boundedness and stability of the solutions of the stochastic flexibility function, given in (\ref{eq:1})-(\ref{eq:4y}). 

\textbf{Theorem 4}\label{thm4}
    \textit{Let $(X^*,u^*)\in \Omega_s$ be an equilibrium point of the stochastic flexibility function (\ref{eq:1}-\ref{eq:4y}). Then given any initial condition, $X_{t_0}$, in $\mathbb{D}$, and any $B^*\in [0,\ 1]$, $C$, and $\Delta$, and $ u^*\in \{0,\ 1 \}$, there exist $\sigma>0$ such that for $\sigma_x\in [0,\ \sigma ]$, the equilibrium points of (\ref{eq:1})-(\ref{eq:4y}) are stochastically bounded.
}

\textbf{Proof}
    Consider the candidate function $V(\mathcal{X}_t)=\frac{1}{2}(\mathcal{X}_t-X^*)^2$
which is $C^2$ on ${\mathbb{D}}\setminus
    {{X}^*}$. It can be shown that there exist positive definite functions $k_1=c_1(\mathcal{X}_t-X^*)^2$ and $k_2=c_2(\mathcal{X}_t-X^*)^2$, with $0<c_1<0.5$ and $0.5<c_2$ such that $k_1\leq V \leq k_2$. Furthermore,
\begin{align}\label{eq:12}
&\mathcal{L}V(\mathcal{X}) = \frac{1}{C}(D_t-B_t)(\mathcal{X}-X^*)+\frac{1}{2}\mathcal{X}^2(1-\mathcal{X})^2\sigma_x^2\notag \\
&=\frac{\Delta}{C} \ell(f(\mathcal{X}_t,\alpha)+g(u_t,\beta),k)(\vmathbb{1}_{\delta_t>0}(1-B_t)+\vmathbb{1}_{\delta_t<0}B_t)
\notag \\
&\times (\mathcal{X}-X^*)+\frac{1}{2}\mathcal{X}^2(1-\mathcal{X})^2\sigma_x^2,
\end{align}
that consists of two terms. The first term is $\eta\ell(f(\mathcal{X}_t,\alpha)+g(u_t,\beta),k)(\mathcal{X}-X^*) $, with $\eta \coloneqq  \frac{\Delta}{C}(\vmathbb{1}_{\delta_t>0}(1-B_t)+\vmathbb{1}_{\delta_t<0}B_t) $, and the second term is $\frac{1}{2}\mathcal{X}^2(1-\mathcal{X})^2\sigma_x^2$. It is noted that the second term is always positive for $\mathcal{X}\in {\mathbb{D}}\setminus
    {{X}^*}$ and $\eta$ is a positive value. In the sequel, we first show that the first term is negative and then complete the proof by finding the conditions for $\mathcal{L}V(\mathcal{X})\leq 0$. 

Since $f(.)$ is monotonically decreasing, $f(\mathcal{X})+g(u^*)>0$ implies $\mathcal{X}<X^*$ and $f(\mathcal{X})+g(u^*)<0$ implies $\mathcal{X}>X^*$, for $(X^*,u^*)\in \Omega_s$. In addition, $\ell$ is strictly increasing. Therefore, $\eta\ell(f(\mathcal{X}_t,\alpha)+g(u^*,\beta),k)(\mathcal{X}-X^*) $ is always negative for $\mathcal{X}\in {\mathbb{D}}\setminus
    {{X}^*}$.     

Considering (\ref{eq:12}), it follows
\begin{align}\label{eq:13w}
&\mathcal{L}V(\mathcal{X})\notag \\
&= \eta\ell(f(\mathcal{X}_t,\alpha)+g(u^*,\beta),k)(\mathcal{X}-X^*)+\frac{1}{2}\mathcal{X}^2(1-\mathcal{X})^2\sigma_x^2\notag \\
&= -\eta|\ell(f(\mathcal{X}_t,\alpha)+g(u^*,\beta),k)||(\mathcal{X}-X^*)|+\frac{1}{2}\mathcal{X}^2(1-\mathcal{X})^2\sigma_x^2 \notag \\
&\leq -\eta_1|\ell(f(\mathcal{X}_t,\alpha)+g(u^*,\beta),k)||(\mathcal{X}-X^*)|+\frac{\sigma_x^2}{32},
\end{align}
where $\sigma_x^2/32$ is the maximum value of $\frac{1}{2}\mathcal{X}^2(1-\mathcal{X})^2\sigma_x^2$ and $\eta_1=\frac{\Delta}{C}\min\{B^*,\ 1-B^* \}$. Then, $\mathcal{L}V(\mathcal{X})\leq 0$ if
 \begin{align}
\frac{\sigma_x^2}{32\eta_1}\leq |\ell(f(\mathcal{X}_t,\alpha)+g(u^*,\beta),k)||(\mathcal{X}-X^*)|\leq |(\mathcal{X}-X^*)|.\notag
\end{align}
This completes the proof for the stochastic boundedness of the trajectories of the flexibility function.
\qed

\textbf{Theorem 5}\label{thm5}
    \textit{Let $(X^*,u^*)\in \Omega_s$ be an equilibrium point of the stochastic flexibility function (\ref{eq:1}-\ref{eq:4y}). Then given any initial condition, $X_{t_0}$, in $\mathbb{D}$, and any $B^*\in [0,\ 1]$, $C$, and $\Delta$, and $ u^*\in \{0,\ 1 \}$, there exist $\sigma>0$ such that for $\sigma_x\in [0,\ \sigma ]$, the equilibrium points of (\ref{eq:1})-(\ref{eq:4y}) are stochastically stable.
}

\textbf{Proof}
   There are two possibilities for the equilibrium points. 
   For $u^*=0$ and $\mathcal{X}\in {\mathbb{D}}\setminus \{1 \}$, $\mathcal{L}V(\mathcal{X})$ is written as
\begin{align}\label{eq:13e}
&\mathcal{L}V(\mathcal{X})
= \eta\ell(f(\mathcal{X},\alpha)+g(0,\beta),k)(\mathcal{X}-1)+\frac{1}{2}\mathcal{X}^2(1-\mathcal{X})^2\sigma_x^2\notag \\
&=|1-\mathcal{X}| \left( -\eta |\ell(f(\mathcal{X}_t,\alpha)+1)|+\frac{1}{2}\mathcal{X}^2|1-\mathcal{X}|\sigma_x^2\right)\notag \\
&\leq |1-\mathcal{X}| \left( -\eta_1 |\ell(f(\mathcal{X}_t,\alpha)+1)|+\frac{1}{2}|1-\mathcal{X}|\sigma_x^2\right).
\end{align}
It can be implied that $\mathcal{L}V(\mathcal{X})\leq 0$ if $-\eta_1 |\ell(f(\mathcal{X}_t,\alpha)+1)|+\frac{1}{2}|1-\mathcal{X}|\sigma_x^2\leq 0$. This inequality can be rewritten as
\begin{align}\label{eq:14e}
-\eta_1 (1-\theta) |\ell(f(\mathcal{X}_t,\alpha)+1)|&-\eta_1 \theta |\ell(f(\mathcal{X}_t,\alpha)+1)|\notag \\
&+\frac{1}{2}|1-\mathcal{X}|\sigma_x^2\leq 0,
\end{align}
where $0<\theta<1$. If $\frac{1}{2}|1-\mathcal{X}|\sigma_x^2\leq \eta_1 \theta |\ell(f(\mathcal{X}_t,\alpha)+1)|$,
then $\mathcal{L}V(\mathcal{X})\leq 0$. Since $\eta_1 \theta |\ell(f(\mathcal{X}_t,\alpha)+1)|\leq \eta_1 \theta$, then $\mathcal{L}V(\mathcal{X})\leq 0$ for $|1-\mathcal{X}|\leq (2\eta_1\theta)/\sigma_X^2$.

Similarly, for $u^*=1$ and $\mathcal{X}\in {\mathbb{D}}\setminus \{0 \}$,
\begin{align}\label{eq:16e}
\mathcal{L}V(\mathcal{X})
&= \eta\ell(f(\mathcal{X}_t,\alpha)+g(1,\beta),k)\mathcal{X}+\frac{1}{2}\mathcal{X}^2(1-\mathcal{X})^2\sigma_x^2\notag \\
&=|\mathcal{X}| \left( -\eta |\ell(f(\mathcal{X}_t,\alpha)-1)|+\frac{1}{2}|\mathcal{X}|(1-\mathcal{X})^2\sigma_x^2\right)\notag \\
&\leq |\mathcal{X}| \left( -\eta_1 |\ell(f(\mathcal{X}_t,\alpha)-1)|+\frac{1}{2}|\mathcal{X}|\sigma_x^2\right).
\end{align}
following the similar discussion for the equilibrium point $(1,\ 0)$, $\mathcal{L}V(\mathcal{X})\leq 0$ for $|\mathcal{X}|\leq (2\eta_1\theta)/\sigma_X^2$.
\qed

\begin{figure}
\label{fig:sim3}
\centering
\includegraphics[width=0.52\textwidth]{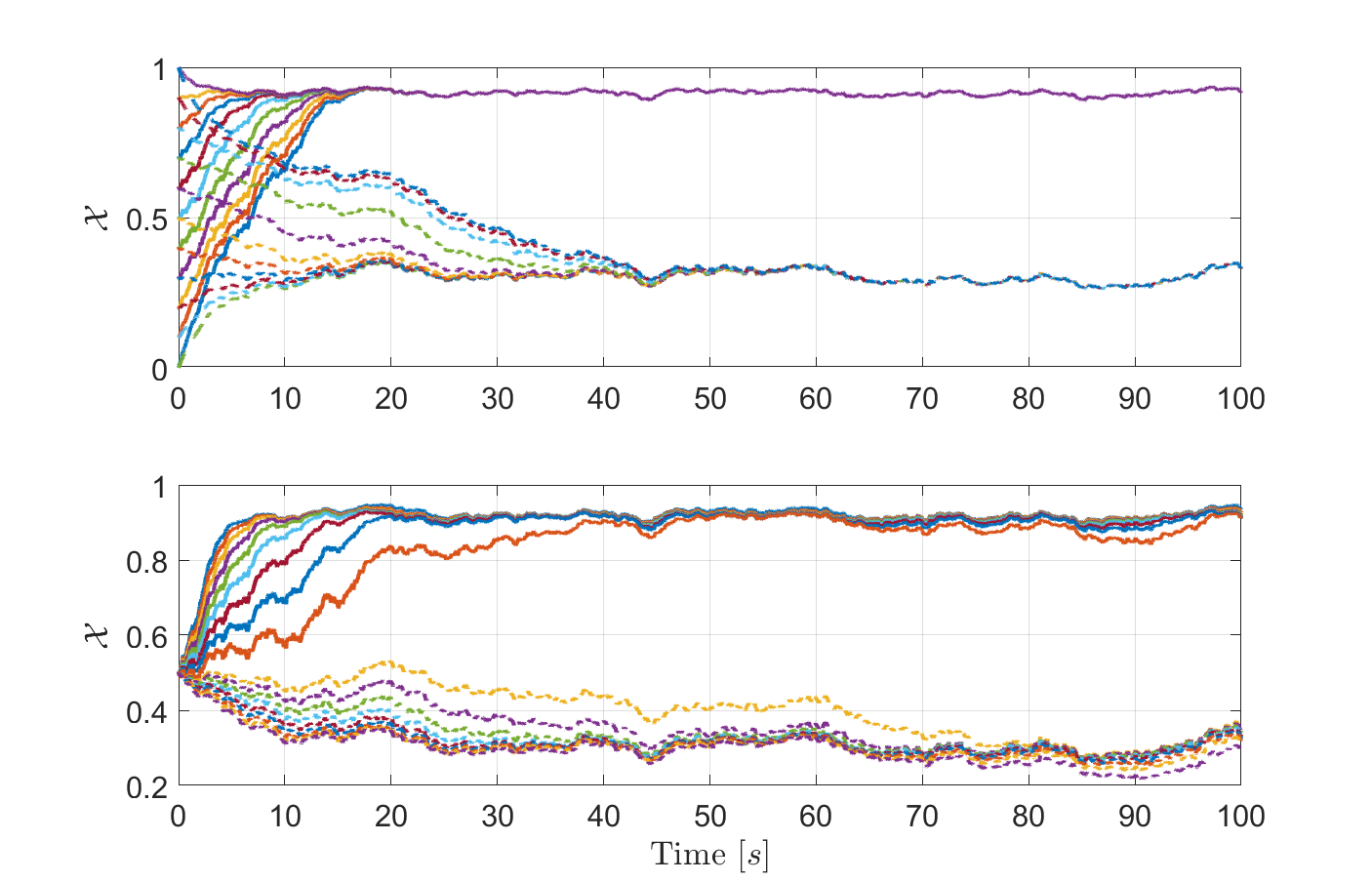}
\caption{State trajectories of the flexibility function with different initial conditions (top panel) and different baseline (bottom panel). state trajectories affected by $u^*=0$ and $u^*=1$ are depicted by solid lines and dashed lines, respectively.}
\end{figure}

\begin{figure}
\label{fig:PDF}
\centering
\includegraphics[width=0.52\textwidth]{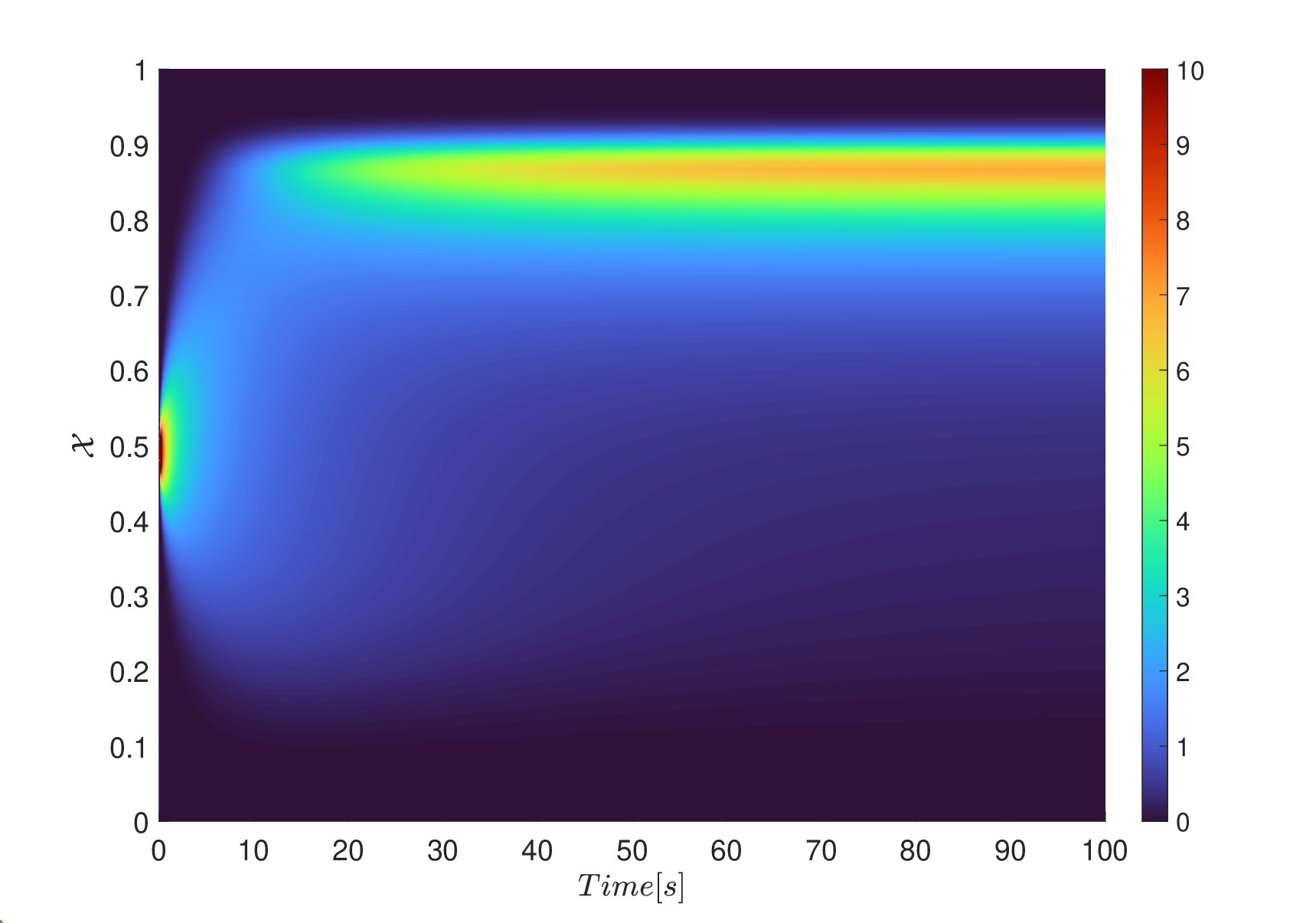}
\caption{Probability density function (pdf) of the state of the flexibility function concerning time.}
\end{figure}

\subsection{Simulation results}

Energy data was carried out using the Energy Zone
Model in IDA-ICE v.4.8 \cite{IDA} for a neighborhood that will be built in Fredrikstad, which is a town located approximately 90 km south of
Oslo, Norway. The total development consists of more than 1500 dwellings, a kindergarten, a school, and commercial buildings \cite{Tohidi22D4.2}. The data was then employed to identify the parameters of the flexibility function.

\begin{figure}
\label{fig:CDF}
\centering
\includegraphics[width=0.52\textwidth]{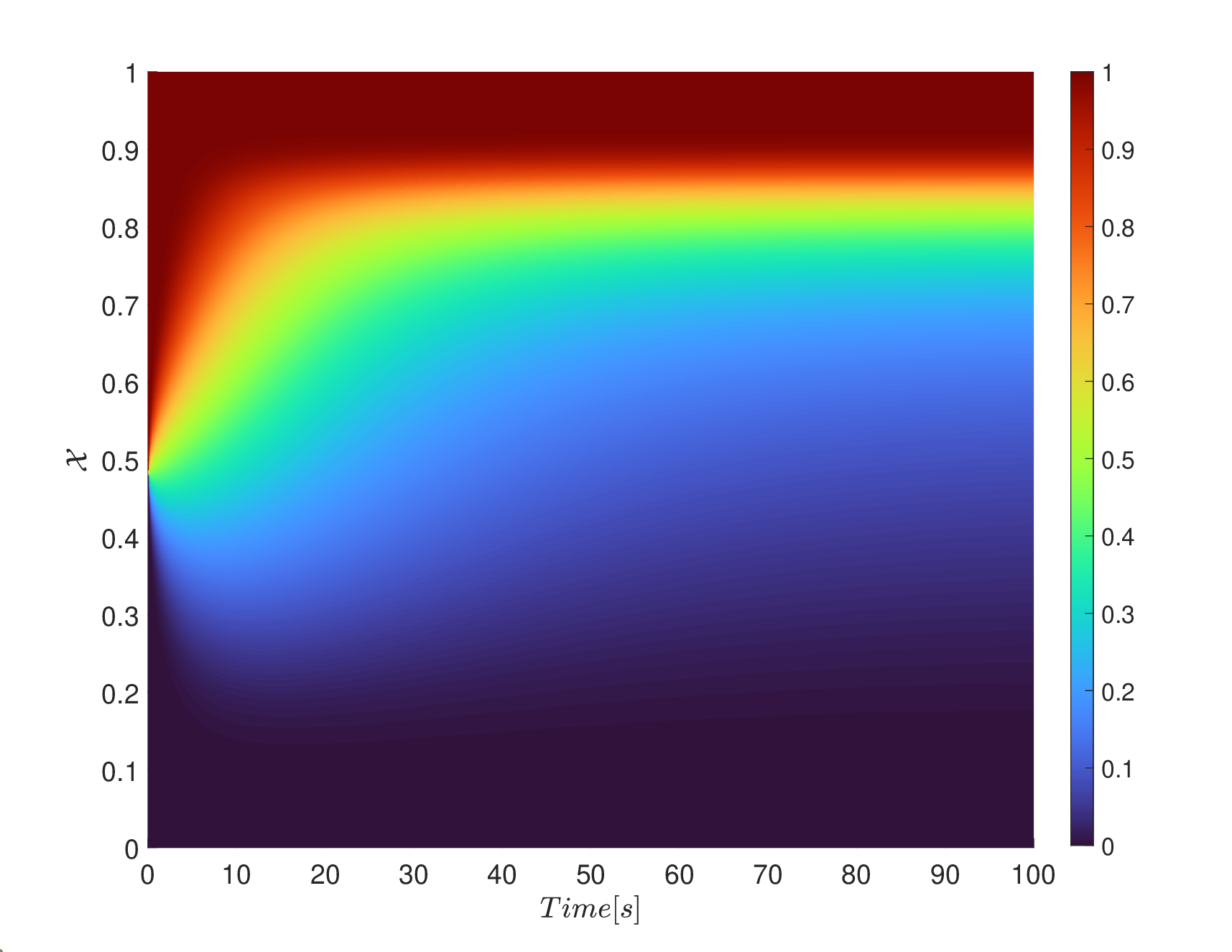}
\caption{Comulative density function (cdf) of the state of the flexibility function with respect to time.}
\end{figure}

\begin{figure}
\label{fig:stat_pdf}
\centering
\includegraphics[width=0.52\textwidth]{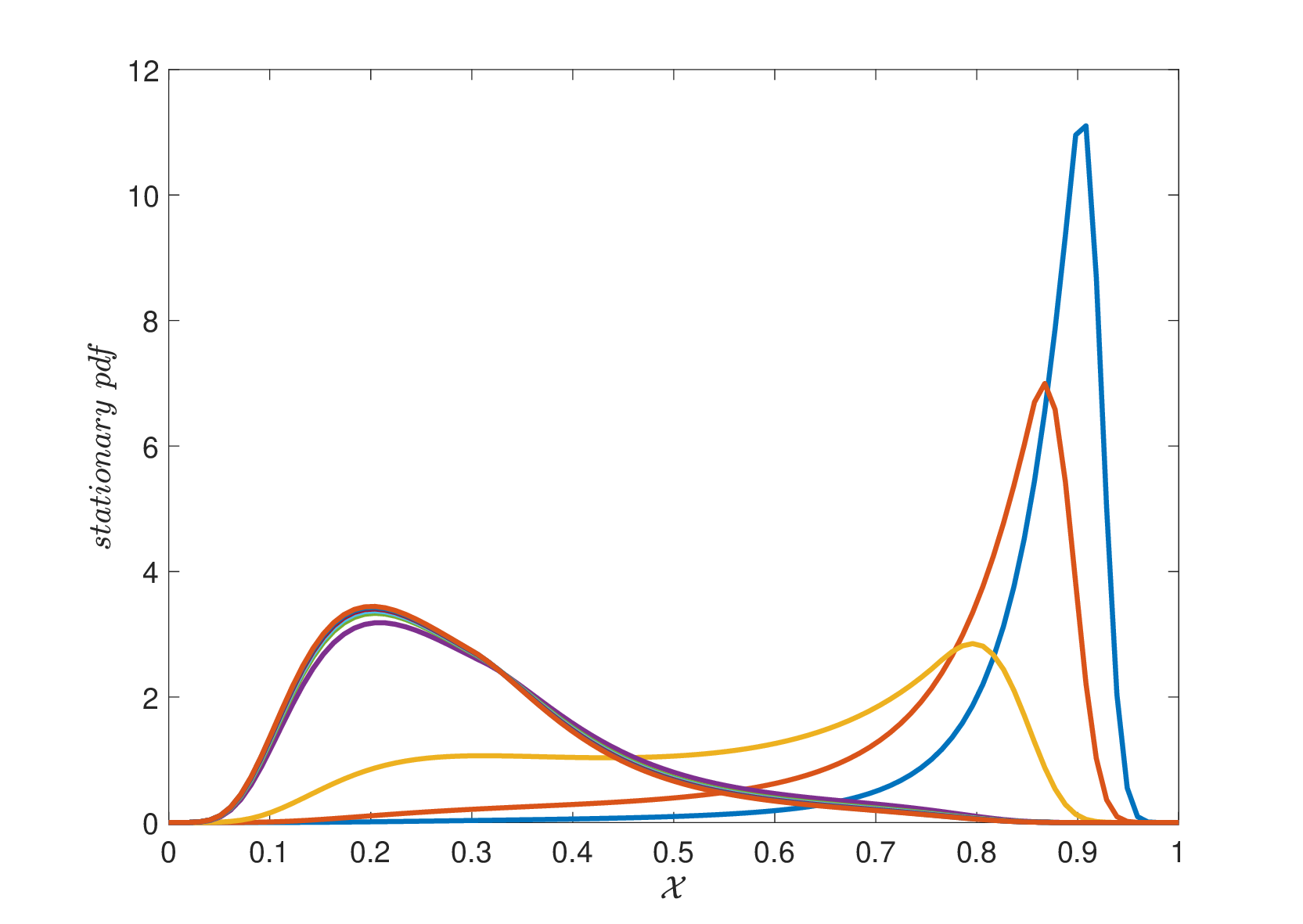}
\caption{Stationary pdf with respect to price change.}
\end{figure}

Figure 4 demonstrates the simulation results of the stochastic flexibility function with parameters $C=2.97$, $\sigma_X=0.1$ and $\Delta=1$. The top panel of this figure shows the state trajectories with different initial conditions, $\mathcal{X}_{t_0}$, and price inputs, $u^*$. The second panel illustrates the state trajectories with different baseline signals, $B^*$, and price inputs, $u^*$. The state trajectories move toward different equilibrium points, $X^*=1$ and $X^*=0$, respectively, and remain in a neighborhood of them. These results follow the stability and boundedness results of Theorems \ref{thm4} and \ref{thm5}.

The probability density function of the state of the flexibility function concerning time is shown in Figure 5. It is seen that with the initial state of $0.5$ and and constant price of $0.2$, the state will converge to a value around $0.86$ with high probability. Figure 6 shows the cumulative distribution function (cdf) of the state with respect to time. At the initial time, $P(\mathcal{X}<0.5)$ is very small. However, as time passes, $P(\mathcal{X}<0.85)$ becomes small. This result is along with the results of pdf in Figure 5.

\begin{figure}
\label{fig:stat_pdf_bchange}
\centering
\includegraphics[width=0.52\textwidth]{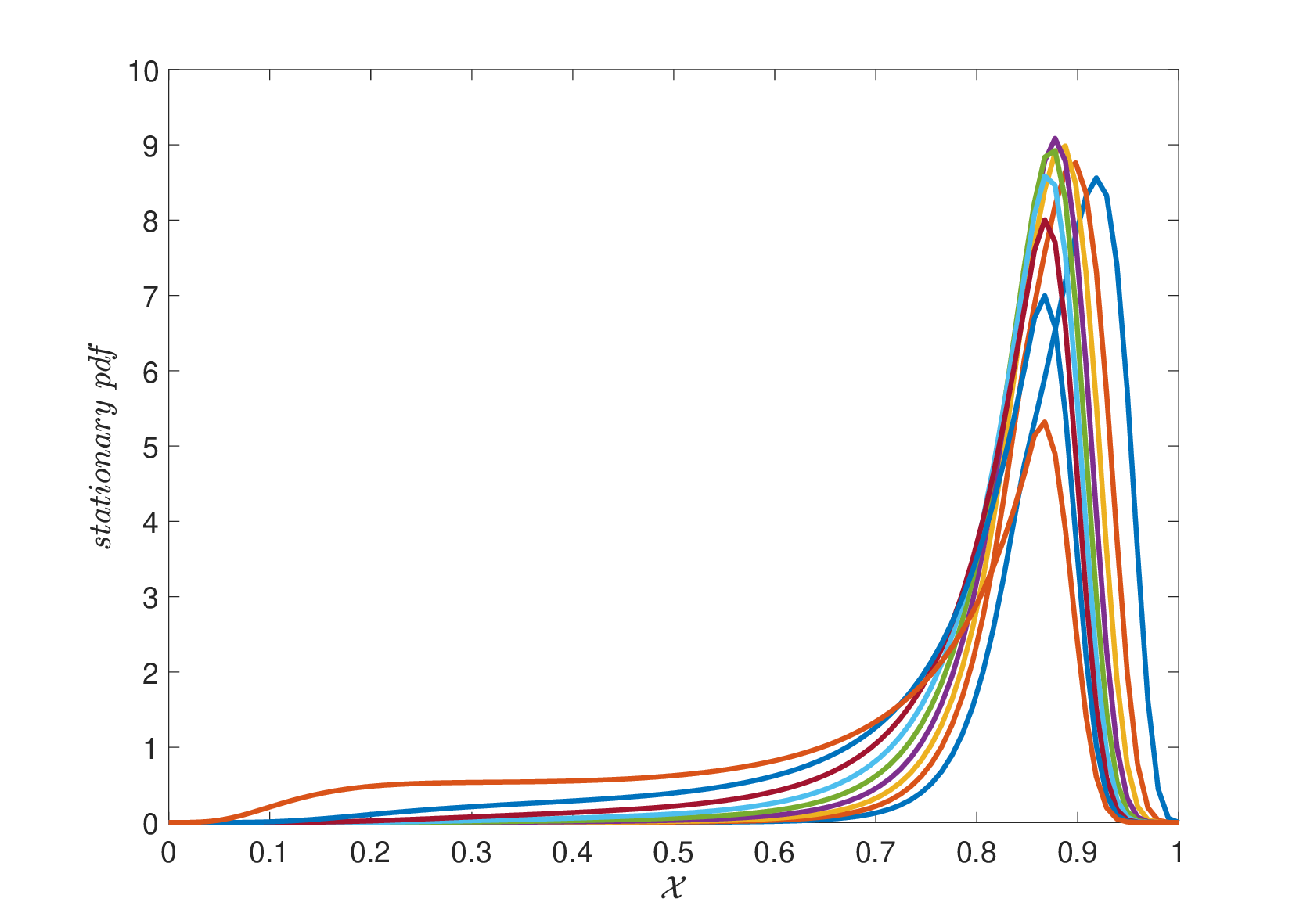}
\caption{Stationary pdf for baseline change.}
\end{figure}

\begin{figure}
\label{fig:sim4}
\centering
\includegraphics[width=0.52\textwidth]{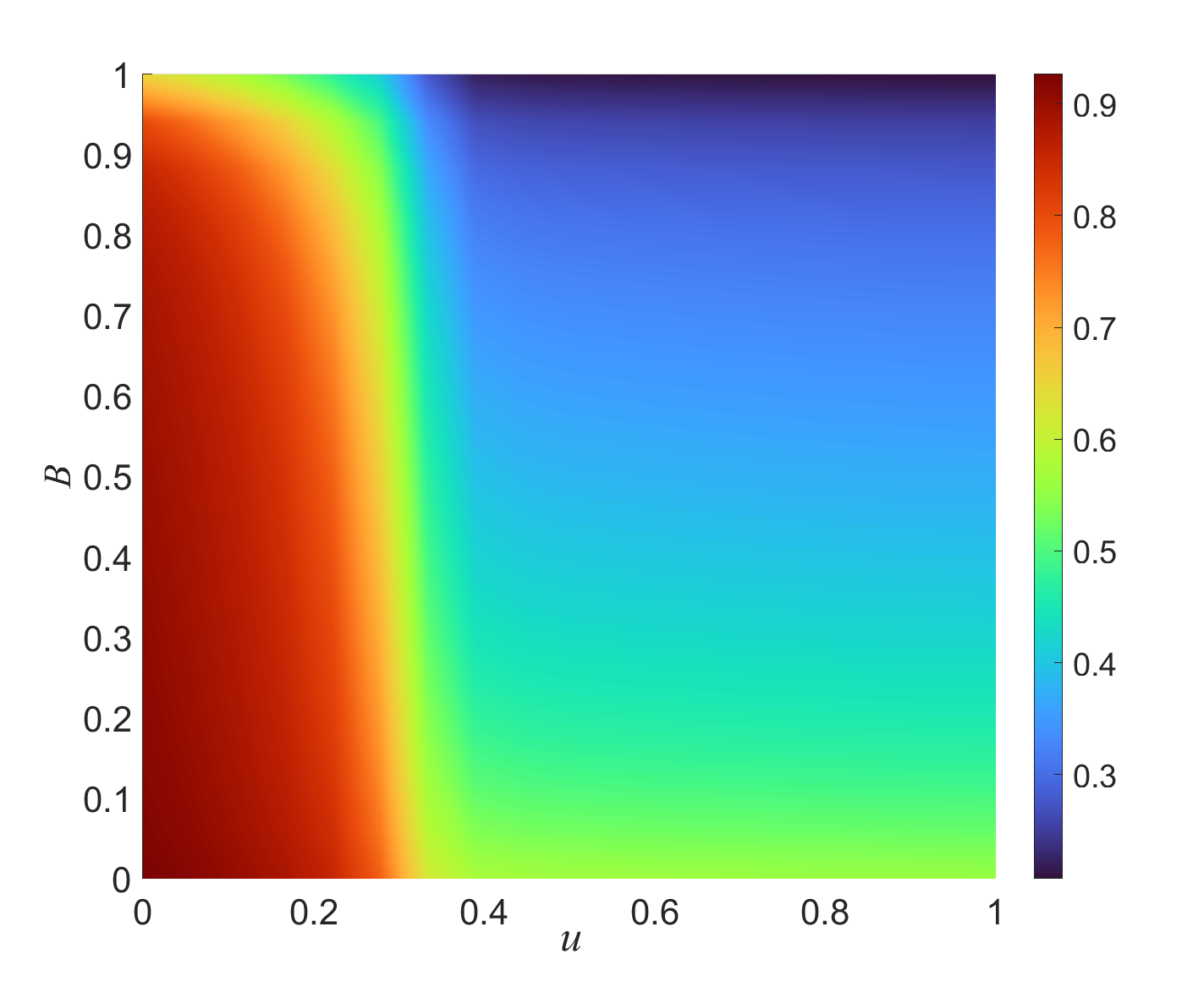}
\caption{Mean of the stationary values of state with respect to different values of $u$ and $B$.}
\end{figure}

Stationary pdf results are provided in Figures 7 and 8. In Figure 7, a stationary pdf is demonstrated concerning various price values. It is seen that for some price values, the state converges to a neighborhood of one, and for others, it converges to a neighborhood of zero. Figure 8 shows that the trajectories converge to a large value (more energy saving), with high probability, once the price is low, i.e. $0.2$, regardless of changing the baseline. 


Figure 9 illustrates the stationary mean values with respect to different values of $u$ and $B$. As expected, low energy price increases the tendency to store more energy and therefore, lead to higher stored energy. Also, at high prices, a high baseline leads to the lowest energy storage. The storage is slightly more when the baseline is low. The variance of the stationary values of the state corresponding to different values of $u$ and $B$ are shown in Figure 10. It is seen that low price leads to a tendency for higher energy storage. The variance is higher for higher energy prices and the highest when the baseline is high while the price is low. Maximum of the absolute value of nonzero eigenvalue of the generator of the Markov chain \cite{Uffe23} with respect to various values of $u$ and $B$ are demonstrated in Figure 11. It is seen that when the price is low, it starts saving energy in a short period of time. When the baseline is high, the energy storage takes place for a longer period of time. Moreover, higher prices lead to some sort of reluctance to change the energy storage in the system. 


\begin{figure}
\label{fig:sim5}
\centering
\includegraphics[width=0.52\textwidth]{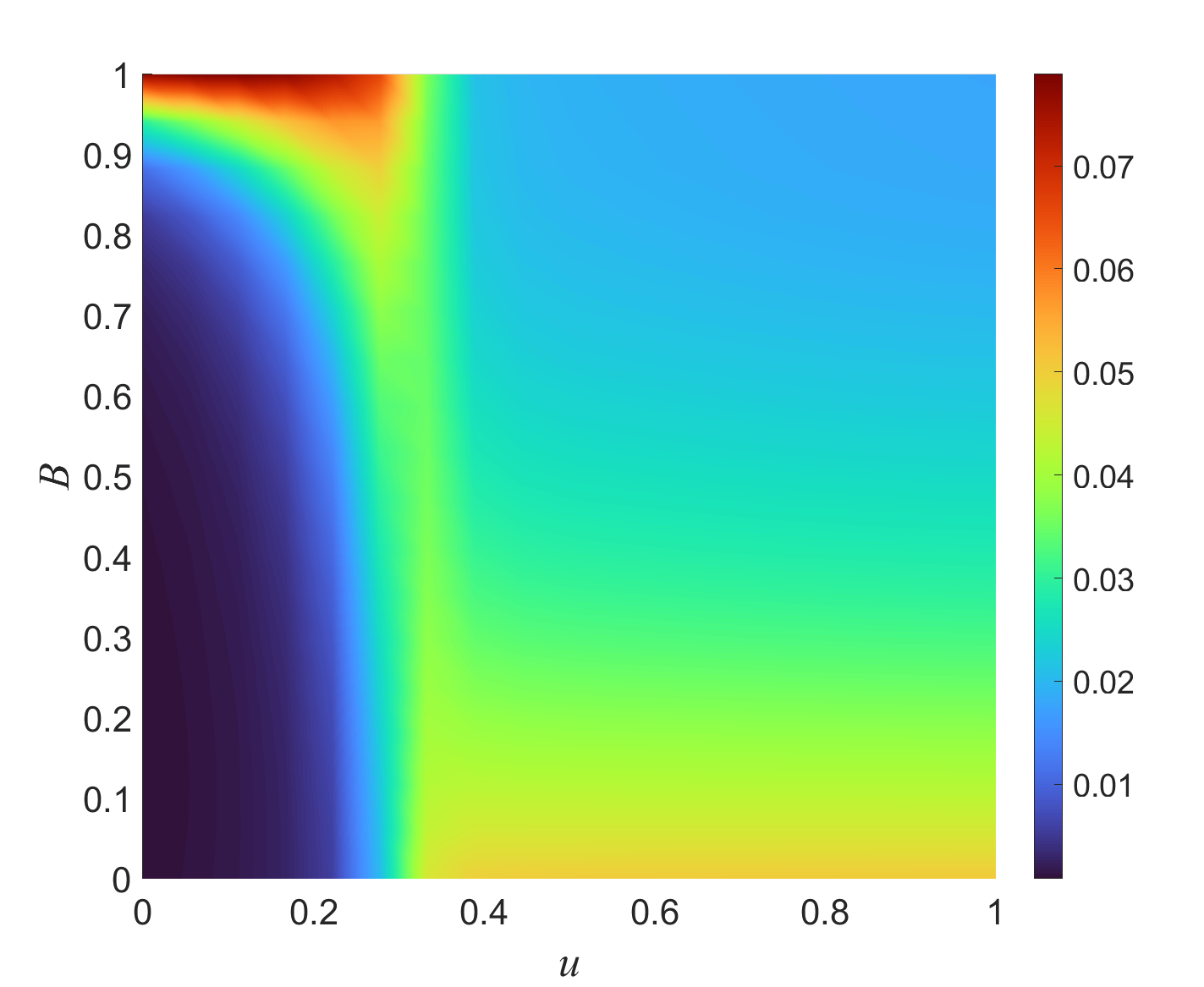}
\caption{Variance of the stationary values of state with respect to various values of $u$ and $B$.}
\end{figure}

\begin{figure}
\label{fig:sim6}
\centering
\includegraphics[width=0.52\textwidth]{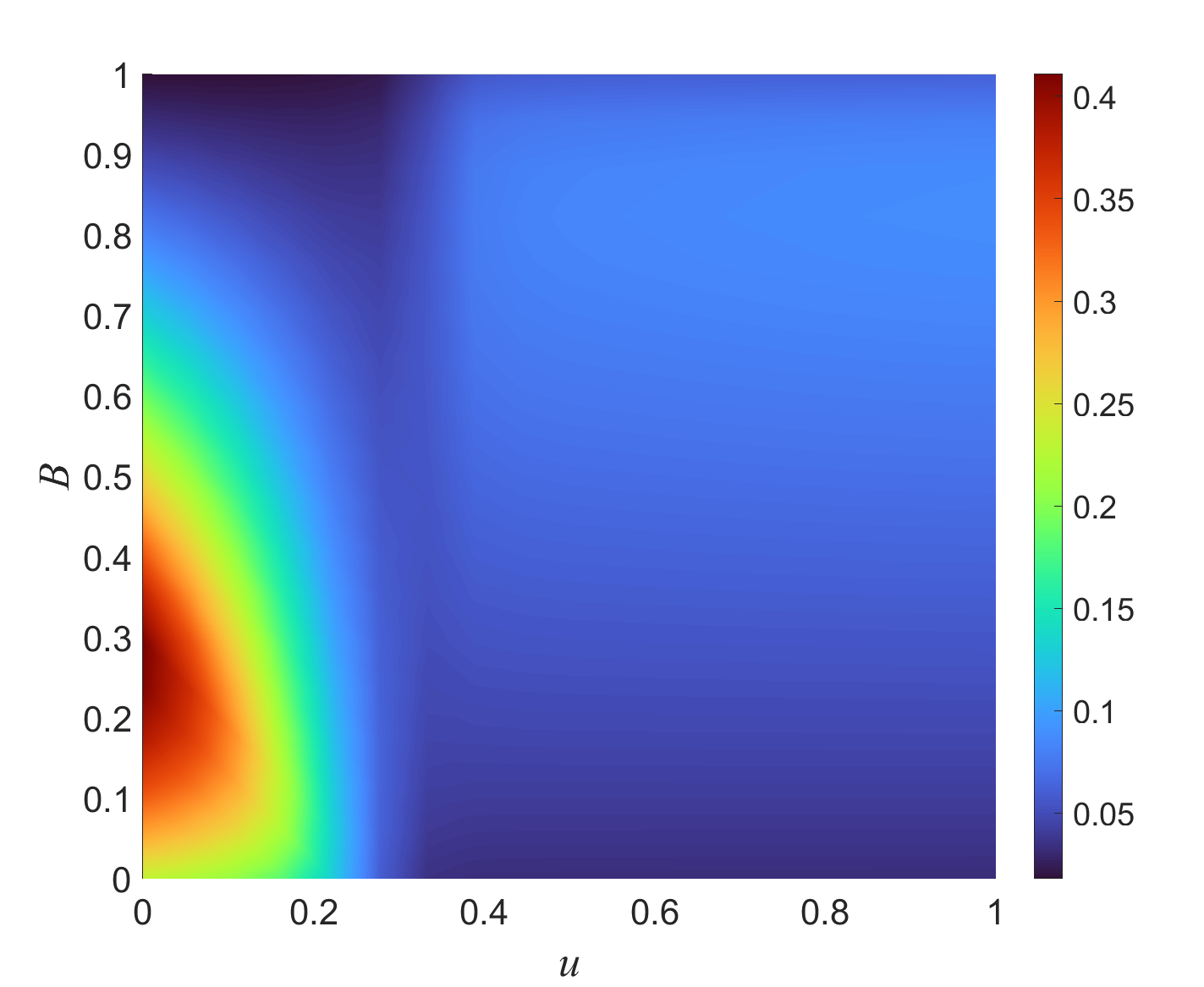}
\caption{Maximum nonzero eigenvalue of the generator of the Markov chain with respect to various values of $u$ and $B$.}
\end{figure}

\section{Conclusions}
In this paper, we analyzed the stability of the flexibility functions, which is an established model of the nonlinear, dynamic, price-demand relationship in energy systems. The stability of the deterministic flexibility function, as a simplified version of the flexibility function, was first investigated. Then, the general stochastic flexibility function is considered and its stability and boundedness are analyzed. Simulation results follow the analysis and illustrate the function's stability properties.

\begin{ack}                               
This work is supported by  \textit{Sustainable plus energy neighbourhoods (syn.ikia)} (H2020 No. 869918), \textit{ELEXIA} (Horizon Europe No. 101075656), \textit{IEA EBC - Annex 81 - Data-Driven Smart Buildings} (EUDP Project No. 64019-0539), \textit{IEA EBC - Annex 82 - Energy Flexible Buildings Towards Resilient Low Carbon Energy Systems} (EUDP Project No. 64020-2131), and the projects \textit{DynFlex} and \textit{PtX, Sector Coupling and LCA}, which both are part of the Danish Mission Green Fuel portfolio of projects.
\end{ack}


\bibliographystyle{unsrt}
\bibliography{autosam}           



\end{document}